\documentclass[twocolumn,showpacs,preprintnumbers,amsmath,amssymb]{revtex4}

\usepackage{graphicx}
\usepackage{dcolumn}
\usepackage{bm}

\begin{document}

\title{Noise correlation spectroscopy of the broken order of a Mott insulating phase}

\author{V. Guarrera}
    \affiliation{LENS European Laboratory for Nonlinear Spectroscopy and Dipartimento di Fisica, Universit\`a  di Firenze,
via Nello Carrara 1, I-50019 Sesto Fiorentino (FI), Italy.}
\author{N. Fabbri}
        \affiliation{LENS European Laboratory for Nonlinear Spectroscopy and Dipartimento di Fisica, Universit\`a di Firenze,
via Nello Carrara 1, I-50019 Sesto Fiorentino (FI), Italy.}
\author{L. Fallani}
        \affiliation{LENS European Laboratory for Nonlinear Spectroscopy and Dipartimento di Fisica, Universit\`a di Firenze,
via Nello Carrara 1, I-50019 Sesto Fiorentino (FI), Italy.}
\author{C. Fort}
    \affiliation{LENS European Laboratory for Nonlinear Spectroscopy and Dipartimento di Fisica, Universit\`a di Firenze,
via Nello Carrara 1, I-50019 Sesto Fiorentino (FI), Italy.}
\author{K. M. R. van der Stam}
        \affiliation{LENS European Laboratory for Nonlinear Spectroscopy and Dipartimento di Fisica, Universit\`a di Firenze,
via Nello Carrara 1, I-50019 Sesto Fiorentino (FI), Italy.}
\author{M. Inguscio}
        \affiliation{LENS European Laboratory for Nonlinear Spectroscopy and Dipartimento di Fisica, Universit\`a di Firenze,
via Nello Carrara 1, I-50019 Sesto Fiorentino (FI), Italy.}

\date{\today}

\begin{abstract}
We use a two-color lattice to break the homogeneous site
occupation of an atomic Mott Insulator of bosonic $^{87}$Rb. We
detect the disruption of the ordered Mott domains via noise
correlation analysis of the atomic density distribution after
time-of-flight. The appearance of additional correlation peaks
evidences the redistribution of the atoms into a strongly
inhomogeneous insulating state, in quantitative agreement with the
predictions.
\end{abstract}


\maketitle

Ultracold atoms in optical lattices are optimum candidates for
quantum computation \cite{quantumcomputing} and quantum simulation
\cite{feynman,jane} thanks to the possibility to control and
access the atomic state. They have demonstrated great capabilities
in reproducing ideal solid state systems, a remarkable example
being the realization of the Mott Insulator (MI) phase
\cite{jaksch98,greiner02}. Furthermore, experiments are now
beginning to access a rather unexplored class of quantum systems
involving correlations and disorder \cite{fallani07}. The need for
effective diagnostic techniques being able to observe these
quantum phases and compellingly establish their nature has become
notable. Few years ago Altman \textit{et al.} suggested in ref.
\cite{altman04} the use of Hanbury Brown and Twiss (HBT) spatial
interferometry to probe hidden order in strongly correlated phases
of ultracold atoms. This proposal has been recently employed to
experimentally investigate quantum statistically correlated
lattice systems \cite{foelling05,rom06,spielman07} and momentum
pair-correlated fermions after molecule dissociation
\cite{greiner05}. Meanwhile, theoretical works have extended the
use of noise interferometry to study disordered systems
\cite{rey06,roscilde07} and to observe a predicted supersolid
phase \cite{scarola06}.

In this work we break the ordered domains of a one-dimensional MI
in a controlled way, by means of a second periodic potential
superimposed on the main lattice. The MI is a state of matter in
which strong interactions force bosonic particles to localize at
the lattice sites, in the homogenous case with exactly the same
number of particles per site. This energetic stiffness of the
phase remains even if, as it happens in usual experiments with
cold atoms in optical lattices, a "mildly" varying trapping
potential is superimposed. This variation of site energy produces
the well known "wedding cake" structure \cite{foelling06}, in
which the system rearranges regularly alternating MI domains with
different occupation number and superfluid islands. Locally the
main properties of the homogeneous system are still preserved,
and, as confirmed by experiments \cite{greiner02,gerbier05} the
global system has mostly an ordered insulating character. A more
intimate alteration of the MI structure can be instead obtained by
a non-monotone scrambling of the energy offsets of the sites,
employing a two-color lattice.

As earlier shown in ref. \cite{fallani07,guarrera07}, the
superposition of an additional weak lattice to a MI state is
enough to profoundly affect the excitation spectrum of the system.
When this second lattice is strong enough, we expect the atoms to
redistribute over the lattice breaking the ordered structure of
the MI. In this Letter we describe an experiment where we have
detected this redistribution using the noise interferometry
technique in a yet unexploited way. In the experiments performed
until now, the noise correlation has only given information on the
strongly correlated character of the MI and on the spatial
ordering of the in-trap density distribution, namely the fact that
atoms are located at equidistant positions in space. In this work
we further demonstrate the powerfulness of this diagnostic
technique by showing that it can be used to detect modifications
in the ordered site filling typical of a MI state, whereas the
density distribution after time-of-flight (TOF) does not show
qualitative differences.
\begin{figure}[t!]
\begin{center}
\includegraphics[width=0.4\textwidth]{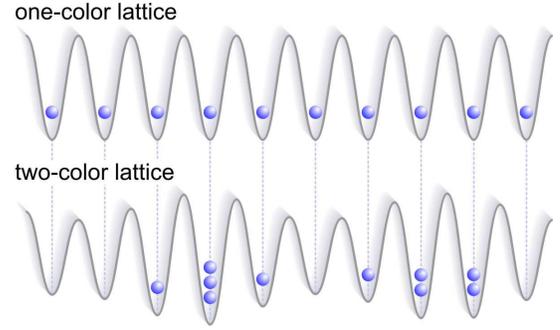}
\end{center}
\caption{In a MI domain the atoms are homogeneously distributed
pinned by strong interactions at the lattice wells. The
superposition of a second weaker lattice with different color
(shown in the lower row) breaks the translational invariance
leading to a redistribution of particles among lattice sites
resulting in a final non-homogeneous density configuration. The
change in the site occupation is the main effect of the additional
lattice, as the position of the wells has not varied
significantly.} \label{fig:atom distribution}
\end{figure}

In the experiment we load a $^{87}$Rb Bose-Einstein condensate
(BEC) containing typically $3 \times 10^{5}$ atoms in a 2D optical
lattice (\textit{confining lattice}) with height $s_{t}=40$, in
units of the recoil energy $E_{R1}=\hbar^{2}k_{1}^{2}/2m \simeq h
\times 3.33$ kHz, where $2\pi/k_{1} = 830.3(1)$ nm is the
wavelength of the lattice light we derive from a Titanium-Sapphire
laser and $m$ is the atomic mass. In this way we create a
bidimensional array of 1D atomic tubes with maximum $200$ atoms,
among which the tunneling is highly suppressed on the timescale of
the experiment. Simultaneously, we apply along the axis of each
atomic tube a third optical lattice (\textit{main lattice}) with
the same spacing and height $s_{1}=16$.

The system can be described by the Bose-Hubbard Hamiltonian
\cite{jaksch98}
\begin{equation}
\hat{H}= -J\sum_{\langle
i,j\rangle}\hat{a}_{i}^{\dag}\hat{a}_{j}+\frac{1}{2}U\sum_{j}\hat{n}_{j}(\hat{n}_{j}-1)+\sum_{j}\epsilon_{j}\hat{n}_{j}
\label{eq:bose-hubbard hamiltonian}
\end{equation}
where $\hat{a}_{j}$ and $\hat{a}_{j}^{\dag}$ are respectively the
annihilation and creation operators of one boson in the $j$-th
site of the lattice and $\hat{n}_{j}$ is the number operator. The
relevant energy scales in the process are the tunneling energy
$J$, the on-site repulsive atomic interaction $U$ and the site
energies $\epsilon_{j}$ determined by the external trapping
potential. In our realization, the system is deeply in the MI
phase: the interaction energy is $U=h \times 2.7$ kHz and the
tunneling along the tubes direction (that we will call
$\textbf{x}$ direction in the following) is $J=h \times 20$ Hz.
The lattice beams have a gaussian shape with a typical waist of
$\sim 200$ $\mu$m and cause an additional external harmonic
confinement. The resulting trapping frequencies are $\omega_x= 2
\pi \times 77$ Hz along the direction of the tubes and
$\omega_{\perp}= 2 \pi \times 110$ Hz orthogonally.

By adding a weaker second lattice with incommensurate wavelength
$2\pi/k_{2}=1076.9(1)$ nm with respect to the main one, along the
same direction, we manipulate the third term in the Hamiltonian
(\ref{eq:bose-hubbard hamiltonian}) according to:
\begin{equation}
\epsilon_{j} = s_{2}E_{R2}\sin^{2} \left( j \pi \frac{k_{2}}{k_{1}}\right) + \frac{m}{2} \omega_x^{2} \left( j\frac{\pi}{k_{1}}\right) ^{2}
\label{eq:e_j}
\end{equation}
with $s_{2}$ the height of the second lattice expressed in units
of the recoil energy $E_{R2}=(\hbar k_{2})^{2}/(2 m)\simeq h
\times 1.98$  kHz.
This additional lattice has the main effect of
producing a non-periodic modulation of the energy minima over a
length scale of $1.8$ $\mu$m, thus breaking the translational
invariance of the main lattice. As far as $s_{2}\ll s_{1}$, the
parameters $J$ and $U$ and the spatial position of the energy
minima $x_{j}$ are only slightly affected by the additional
lattice \cite{guarrera07}. Increasing $s_{2}$, the ordered atomic
distribution in the Mott domains is gradually affected and atoms
are expected to rearrange in a non uniform pattern as shown in
Fig. \ref{fig:atom distribution}.
\begin{figure}[t!]
\begin{center}
\includegraphics[width=0.3\textwidth]{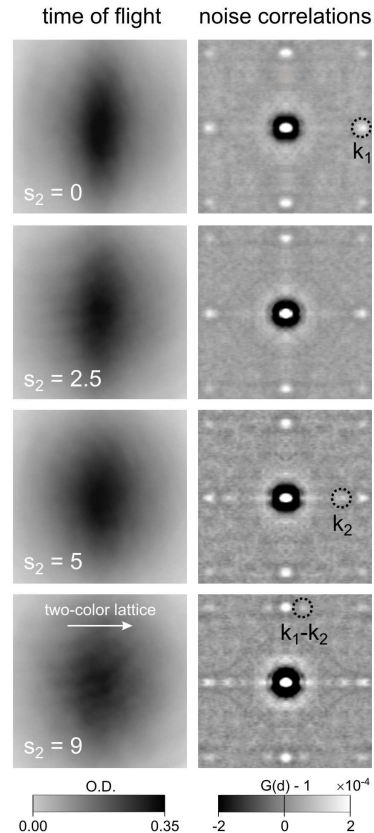}
\end{center}
\caption{Left column: averaged images of the atomic
density distribution after $20$ ms time-of-flight with $s_{1}=16$ and varying
$s_{2}$. Right column: correlation functions. For $s_2
\lesssim 3$ peaks at a distance $d=2 \hbar k_1 t/m$ appear
reflecting the main lattice periodicity. When $s_{2}>5$ we observe
the growth of additional peaks corresponding to the periodicity of
the second lattice ($k_{2}$) and to the
beating between the two standing waves ($k_{1}-k_{2}$).}
\label{fig:signal}
\end{figure}

We probe the new density distribution using the noise correlation
analysis at fixed height of the main lattice $s_{1}=16$ and
various second lattice intensities in the range $0\leq s_{2} \leq
9$. We load the BEC into the final optical lattice configuration,
switching on all the lattices at the same time by means of an
exponential ramp with time constant $30$ ms and duration $140$ ms.
After holding the atoms for $30$ ms in the lattices, we let the
atomic system expand by suddenly switching off all the confining
potentials. After $20$ ms of ballistic expansion, we image the
cloud by standard destructive absorption technique. The probe beam
has an intensity $I=0.5$ mW/cm$^{2} \simeq 0.3I_{sat}$ where
$I_{sat}$ is the saturation intensity of the $D_{2}$ transition.
We illuminate the atomic cloud for $50$ $\mu$s and then we image
the transmitted light on an interline CCD camera.

From the obtained images we calculate the normalized
density-density correlation function analogously to what is done
in ref. \cite{foelling05}, by:
\begin{equation}
\mathcal{G}(\textbf{d})= \frac{\int d \textbf{r} \langle \hat{n}(\textbf{r}) \hat{n}(\textbf{r}+\textbf{d})\rangle_{t} }{\int d \textbf{r} \langle \hat{n}(\textbf{r}) \rangle_{t} \langle \hat{n}(\textbf{r}+\textbf{d})\rangle_{t}}.
\label{eq:correlation function}
\end{equation}
where $\hat{n}(\textbf{r})$ is the column density detected at the
position $\textbf{r}$ reached by the particles in a time $t$ after
releasing from the trap, $\textbf{d}$ is the separation between
two detected positions in the expanded atomic cloud and $\langle
\rangle$ refers to a statistical averaging over a set of several
images taken in the same conditions \cite{fft}. A Butterworth
high-pass filter is applied on the correlation signal to eliminate
the gaussian background appearing as a consequence of shot-to-shot
$\pm 25 \%$ fluctuations in total atom number. A significative
increase in the visibility of the correlation peaks has been
obtained by averaging the final image with itself after flipping
around the central vertical axis. This procedure is justified by
the symmetry of the lattice system under inversion $x \rightarrow
-x$.

Non-trivial correlations have shown to be present in the
second-order correlation function $\mathcal{G}(\textbf{d})$ of the
density distribution of a bosonic atomic cloud freely expanding
from a MI state, with peaks emerging spaced by the distance
$2\hbar k t /m$ \cite{altman04,foelling05}, being $k$ the
wavevector of the lattice laser light. Figure \ref{fig:signal}
shows images of the density distribution after TOF (left column)
and the correlation function (right column) both averaged on
$5\div 6$ sets of $40\div 80$ images for different values of
$s_{2}$. The TOF images exhibit an unstructured gaussian profile
and no significative differences increasing the height of the
second lattice (apart from an increase in size along the direction
of the two-color lattice). When $s_{2} \lesssim 3$ the measured
correlation signal presents peaks clearly visible in
correspondence of the periodicity of the main lattice ($k_{1}$
peaks) and no significative alteration with respect to the pure MI
case. At $s_{2}=5$ additional peaks are detectable at the
periodicity of the second lattice ($k_{2}$ peaks), in the central
row of the noise correlation signal. Similar correlation peaks
have been predicted to appear first for hard-core bosons in a
quasiperiodic potential \cite{rey06} and then also for soft-core
bosons \cite{roscilde07}. The additional peaks strengthen,
becoming clearly visible at $s_{2}=9$, where also the beating
between the two standing waves ($k_{1}-k_{2}$ peaks) is visible in
the upper and lower rows. Note, that the additional peaks only
appear along the horizontal direction where the second lattice
modifies the site occupation. The $k_{1}-k_{2}$ peaks should be
visible also in the central row, but they are obscured by a
central black ring artifact deriving from the application of the
high-pass filter.
\begin{figure}[t!]
\begin{center}
\includegraphics[width=0.5\textwidth]{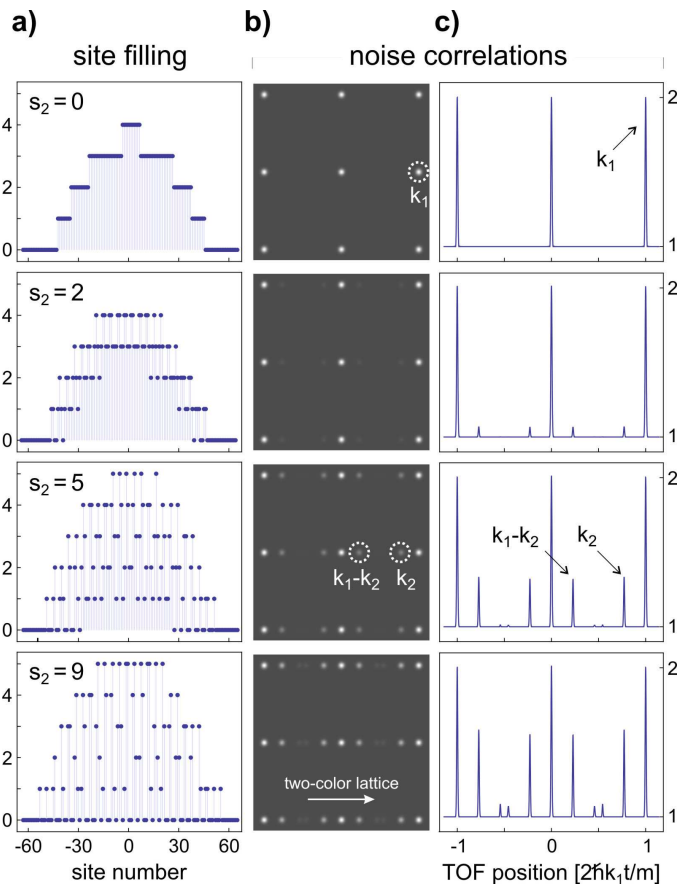}
\end{center}
\caption{The ground state of the system is calculated by
diagonalizing Eq. \ref{eq:bose-hubbard hamiltonian} in the limit
$J\longrightarrow 0$. (a)
Calculated atom density distribution at $s_{1}=16$ and
$s_{2}=0,2,5,9$. (b) Noise correlation signal computed for
each of the ground states given in (a). (c) Horizontal
cross section of the peaks in the noise correlation signal.}
\label{fig:theoretical model}
\end{figure}
Because the second lattice does not appreciably shift the position
of the potential wells, in the range of parameters we used, the
appearance of the additional peaks arises from a different effect,
namely the change in the local atom site occupation induced by the
second lattice. Atoms rearrange in the main lattice wells
searching for the lowest energy configuration driven by the
interplay between the second-color modified site energies and the
interactions.

In order to obtain a deeper insight, we have developed a
theoretical model working in the zero-tunneling limit which
confirms this interpretation. In this limit the Hamiltonian in
equation (\ref{eq:bose-hubbard hamiltonian}) is diagonal on the
tensor product of number Fock states in each lattice site,
describing a MI state. The ground state is given by the
distribution of atoms minimizing the total energy of the system.
The lowest energy configuration is computed, fixing the number of
atoms, and subsequently noise correlations are calculated
according to ref. \cite{altman04}. In Fig. \ref{fig:theoretical
model} we report the result of this calculation for different
values of the second lattice intensity $s_{2}=0,2,5,9$. The
picture shows in column: (a) the density distribution for the
central atomic tube, (b) the 2D correlation signal and (c) its
cross section along the horizontal line. At $s_{2}=0$ the system
shows the typical wedding cake structure, due to the presence of
the harmonic trapping potential, with MI domains of $N=1,2,3,4$
atoms per site. The corresponding noise signal exhibits the peaks
at periodicity $k_{1}$ as expected. When $s_{2}=2$ the MI order
begins to be destroyed starting from the edge of each domain,
where the site energies scrambling is more efficient due to the
different atom occupation of next-neighboring sites with almost
the same energy. The noise signal shows weak peaks emerging at the
periodicity $k_{2}$ and $k_{1}-k_{2}$, as can be seen in the cross
section picture. Increasing the height of the second lattice to
$s_{2}=5$, the density profile gets more and more inhomogeneous,
affecting also the bulk of the domains. The extra peaks are now
clearly visible in the noise signal. At the maximum value
$s_{2}=9$, sites with $N=5$ occupation appear together with empty
sites even at the trap center and the wedding cake structure is
completely lost being washed away in favor of the second-color
driven density distribution. The cross section graph shows the
appearance of higher-order beating peaks. Note that the peaks at
$k_{2}$ are slightly higher than the ones at $k_{1}-k_{2}$. We
have verified in our numerical simulations that this is the only
effect due to the small modification of the minima position
induced by the strong second lattice (the relative deviation with
respect to the main lattice spacing is $\sim 4 \% $ at $s_{2}=9$).

On a qualitative level, the theoretically derived
signals show the same behavior as the experimental observations, see
Fig.\ref{fig:signal}.
\begin{figure}[t!]
\begin{center}
\includegraphics[width=0.5\textwidth]{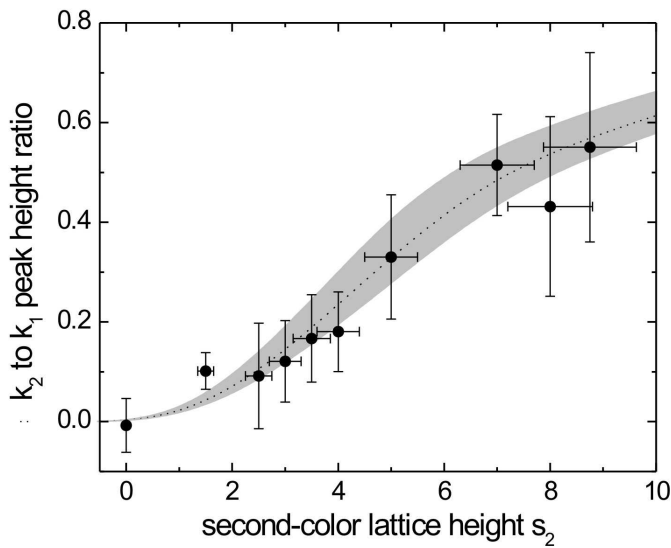}
\end{center}
\caption{The ratio
between the height of the $k_{2}$ peak and the height of the $k_{1}$
peak is reported as a function of $s_{2}$, while $s_{1}$ is fixed at
the value of $16$. Each point is an average of $5\div 9$ values from
different sets of images. The
grey curve derives from calculation by a zero-tunneling model for a
total atom number ranging from $2\times 10^{5}$ to $4 \times
10^{5}$.} \label{fig:peaks ratio}
\end{figure}
For a more quantitative analysis we compare in Fig.\ref{fig:peaks
ratio} the theoretical and the experimental results for the ratio
between the height of the $k_{2}$ peak and the main $k_{1}$ peak
for different values of $s_{2}$. Each point in the graph results
from the weighted average of $5-9$ experimental values. The dotted
curve shows the theoretical prediction for $N=3 \times 10^{5}$
atoms. The grey zone takes into account possible systematic errors
in the measurement of the absolute number of atoms, showing the
results of the calculations for a number of atoms ranging from $2
\times 10^{5}$ to $4 \times 10^{5}$. We find good agreement
between the experimental data and the results of this
zero-tunneling model with no free parameters. Results compatible
with our experimental findings have also been obtained by T.
Roscilde \cite{roscilde07} with a Quantum Monte Carlo calculation
performed for a one dimensional system with $N=100$ atoms,
corresponding to one atomic tube of our realization.

In conclusion, we have destroyed in a controlled way the Mott
Insulator ordered structure by using an additional lattice at a
different incommensurate wavelength and we have monitored its
progressive degradation via quantum noise interferometry. We have
demonstrated that the noise correlation technique can be employed
to detect modifications in the occupation of the lattice sites.
Furthermore, a zero-tunneling model shows good agreement with the
experimental data. Future perspectives of extending our
theoretical model to keep into account finite tunneling appear
interesting, especially for the search of possible signals that
could give information on characterizing complex quantum phases
like the Bose Glass \cite{gianmarchi88,fisher89,fallani07}.

\begin{acknowledgments}
This work has been funded by the UE Contracts
No.HPRN-CT-2000-00125, MIUR PRIN 2005 and Ente Cassa
di Risparmio di Firenze, DQS EuroQUAM Project, Integrated Project
SCALA.
K.M.R. van der Stam acknowledges the donation of Mrs. Aleo which founded
the PostDoc position at Lens.
We thank E. Altman, E. Demler, M. Modugno, T. Roscilde and D.
van Oosten for fruitful discussions, as well as all the
colleagues of the Quantum Degenerate Gases Group at Lens.
\end{acknowledgments}

\end{document}